\documentclass[preprint,12pt,aps]{elsarticle}
\pdfoutput=1 

\usepackage{amssymb}
\usepackage{lipsum}

\usepackage{subfigure}
\usepackage{dcolumn}
\usepackage{bm}
\usepackage{color}
\usepackage{siunitx}
\usepackage{booktabs}
\usepackage{commath}
\usepackage{multirow}
\usepackage[none]{hyphenat}
\usepackage{natbib}
\usepackage{hyperref}

\begin{document}

\begin{frontmatter}

\title{Molecular structure, electric property, and scintillation and quenching of liquid scintillators}%

\author[a,b,c]{Zhe Wang\corref{cor1}}
\author[a,b,c]{Ye Liang}
\author[a,b,c]{and Haozhe Sun}


\affiliation[a]{organization={Department of Engineering Physics, Tsinghua University, Beijing 100084, China}}
\affiliation[b]{organization={Center for High Energy Physics, Tsinghua University, Beijing 100084, China}}
\affiliation[c]{organization={Key Laboratory of Particle \& Radiation Imaging (Tsinghua University), Ministry of Education, Beijing 100084, China}}

\cortext[cor1]{correspondence: wangzhe-hep@mail.tsinghua.edu.cn}


\date{\today}

\begin{abstract}
Liquid scintillators are widely used in particle and nuclear physics. Understanding the scintillation and quenching mechanisms is a fundamental issue in designing a high-light-yield liquid scintillator. 
In this work, the basic scintillation process for two-component liquid scintillators is discussed, highlighting the processes of excitation, ionization, and anion-cation recombination.
A molecule's polar group, polarization characteristics, and the corresponding material's dielectric constant are found to be correlated with a liquid scintillator's scintillation efficiency.
Polar groups and high relative dielectric constant (permittivity) can cause quenching and should be avoided.
The tellurium loading scheme in the liquid scintillator of the SNO+ experiment, TeBD, is discussed. 
The hydroxyl groups introduce polar structures in the TeBD, and for the first time, the relative dielectric constant of TeBD is measured to be $16\pm1$.
These discussions explain part of the quenching of the TeBD liquid scintillator.
\end{abstract}


\end{frontmatter}

\section{Introduction and Motivation}
Liquid scintillators are widely used in particle and nuclear physics. They are also powerful research tools in chemistry, biology, medicine, etc.
For various purposes, to develop a high-light-yield liquid scintillator with a specific isotope loading is difficult~\cite{SNOPlusLS, Scintillator}. 
For instance, people are testing several tellurium loading schemes for the neutrinoless double beta decay study~\cite{SNOPlusTeBD, TeLS1, TeLS2, TeLS3, TeLS4, Liangye} and are looking for an ideal approach.

For a liquid scintillator design, an understanding of the scintillation and quenching mechanisms is expected.
In particular, the connection between the light yield performance and the structure of the involved molecules needs to be well understood. 
For instance, we have been trying to develop a water-based liquid scintillator~\cite{WaterLS}, expecting a lower budget and a better solubility with inorganic salt~\cite{Liangye}. Phenol and benzyl alcohol have been chosen to be mixed with water, because they both have scintillation groups, benzene structure, and hydrophilic groups, hydroxyl (-OH). But the attempts are not successful.
The various tellurium loading plans~\cite{SNOPlusTeBD, TeLS1, TeLS2, TeLS3, TeLS4, Liangye} also show different light yield performance.
The relative light yield usually drops to about 50\% when Te loading has reached 1-2\% (by weight).
We hope to understand the underlying reason and to accelerate the liquid scintillator development. 

This work discusses part of this critical issue, focusing on the electric property of a liquid scintillator.
The electric polar groups and polarization characteristics of liquid scintillator molecules play an important role in the scintillation and quenching process.
The related dielectric constant, also called permittivity, should be measured and is worth noting.

The paper is presented in the following way.
In Sec.~\ref{sec:scintillation}, the scintillation process is briefly reviewed, 
highlighting the process of energy deposition of charge particles and the process of anion-cation recombination.
In Sec.~\ref{sec:quenching}, the electric property related quenching effect is explained, focusing on the impact of polar groups and dielectric constant on the anion-cation recombination.
In Sec.~\ref{sec:TeBD}, the tellurium loading scheme of the SNO+ experiment~\cite{SNOPlusTeBD} is investigated. The electric property of the synthesis product is measured. 
Section~\ref{sec:discussion} summarizes the main conclusion of the work.

\section{Scintillation process}
\label{sec:scintillation}
\begin{figure}[htpb]
    \centering
    \includegraphics[width=0.6\linewidth]{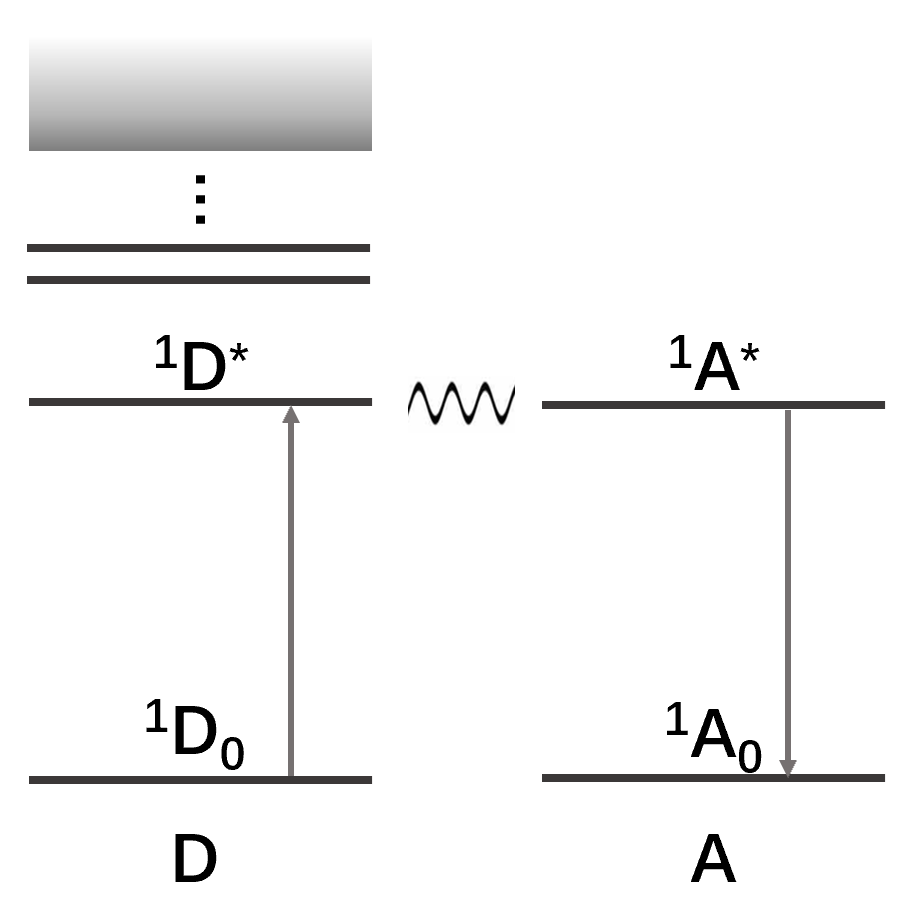}
    \caption{A simplified schematic diagram of the scintillation process. D is for the primary component, donor or solvent, and A is for the secondary component, acceptor or solute. More explanations are in the text.}
    \label{fig:process}
\end{figure}

This paper is about the binary liquid scintillator system, for instance, LAB with 2,5-diphenyloxazole (PPO), in which LAB is the solvent and dominant in mass, and PPO is only at the level of a few grams per liter.
LAB is the primary scintillation component, and PPO is the secondary component. 
The primary component absorbs almost all of the energy deposition of radiation and can emit short-wavelength scintillation light or transfer its excitation state energy to the secondary component. 
The secondary component emits longer wavelength scintillation light for a higher propagation length and for a higher quantum efficiency of an optical photon instrument, for instance, a photomultiplier tube.
The process is presented in Fig.~\ref{fig:process}, where D is for the primary component, donor or solvent, and A is for the secondary component, acceptor or solute.

Elastic scattering, excitation, and ionization are the major physics processes of a charged particle with liquid scintillator molecules.
As presented in Fig.~\ref{fig:process}, solvent molecules are excited from their ground states, ${^1\mathrm{D}_0}$, to their excited states, represented by the horizontal solid lines above, ${^1\mathrm{D}^*}$. 
Here the post-script 0 and * refer to the ground and excited states, respectively. The pre-script, 1, refers to the spin singlet.
If the absorbed energy is very high, electrons are released, and anions and cations are created in the liquid. This transition from the higher and continuous excited states to the quasi-free electrons release is represented by the gray-to-white area in the figure.
Anions (electrons), cations, and excited states are the major products after ionization and excitation processes.

Some of the electrons and cations can recombine into neutral excited states, while some electrons can be solvated in the solvent and finally cannot contribute back to scintillation.
The recombination rate will increase with a higher attraction force between the positive and negative charges and the media, which is governed by Coulomb's Law.
\begin{equation}
    F=\frac{1}{4\pi\epsilon_0\epsilon_r} \frac{q_1 q_2}{r^2},
    \label{eq:Coulomb}
\end{equation}
where $q_1$ and $q_2$ are the charge of an electron and a cation, respectively, $\epsilon_0$ is the dielectric constant or permittivity of vacuum, $\epsilon_r$ is the relative dielectric constant of the media, and r is their distance.
A high recombination force is connected with a low $\epsilon$ and neutral moleculer structures.
On the other hand, 
the electrophilic characteristics of some solvent molecules will facilitate the electron solvation and reduce electron and cation recombination.

The energy is transferred from the solvent molecule's excited states, ${^1\mathrm{D}^*}$, to the second component, A, as in Fig.~\ref{fig:process}, and A is excited from ${^1\mathrm{A}_0}$ to ${^1\mathrm{A}^*}$.
The deexcitation of ${^1\mathrm{A}^*}$ gives the scintillation lights.
The energy transition is conducted dominantly by two mechanisms.
The first one is the F$\mathrm{\ddot{o}}$rster resonance energy transfer (FRET), i.e.,~electric dipole-dipole interaction~\cite{Forster}.
The second one is the charge exchange effect~\cite{Exchange}.
When the concentration of the second component is low, the dipole-dipole process is dominant,
and the transition rate is proportional to the reciprocal of the fourth power of the refractive index, $n$, of the medium.

\section{Electric property related quenching}
\label{sec:quenching}

\subsection{Quenching in anion-cation recombination}
\label{sec:quenchingCombination}

Liquid scintillator molecules' polar groups, their polarization characteristics, and the related dielectric constant of the whole liquid can affect the anion-cation recombination. 
There can be a quantitative connection between the recombination probability and the structures and the relative dielectric constant.

The yield of free ions by radiation has been thoroughly investigated in the past~\cite{FreeIonYield}.
After ionization, usually an electron will return to its parent ion and go back to an excited state, contributing to the scintillation, but some electrons will escape from the initial recombination.
The number of escaped electrons per unit energy input by radiation is called the free ion yield.

The escaped electrons can be quasi-free electrons in a delocalized state in a conduction band, or aggregate with solvent molecules to form a solvated electron, having a much longer lifetime than scintillation.
With a lot of polar groups, a high dielectric constant is granted. 
As expressed in Eq.~\ref{eq:Coulomb}, the recombination attraction force is proportional to the reciprocal of the dielectric constant.

Water has a relative dielectric constant of 78.36 at 0~Hz at 25~$^\circ C$~\cite{CRCHandbook}.
As a comparison, the relative dielectric constants for benzene, toluene, $p$-xylene, ethylbenzene, propylbenzene, and butylbenzene at 0 MHz and 20 $^\circ C$ 
are 2.28, 2.38, 2.27, 2.45, 2.37, and 2.36, respectively~\cite{CRCHandbook}.
The relative dielectric constant of LAB is about between 2.3 to 2.4 as reported later in this paper.

The relative dielectric constant varies with frequency and temperature, since the media's molecules cannot respond instantly to a high-frequency external field.
Higher temperature will induce smaller viscosity and faster response to the external field.
The real part of water's relative dielectric constant is in the range of 78 - 15 for the frequency range of 0~Hz - 50~GHz at 25~$^\circ C$~\cite{CRCHandbook}.
For anion-cation recombination, the lifetime of free electrons can be extremely short and is in the ps range.
Currently, no detailed results for popular organic liquids and water are found for such high frequency, so we chose their values at 0~Hz as a proxy.

\begin{figure}[htpb]
    \centering\includegraphics[width=\linewidth]{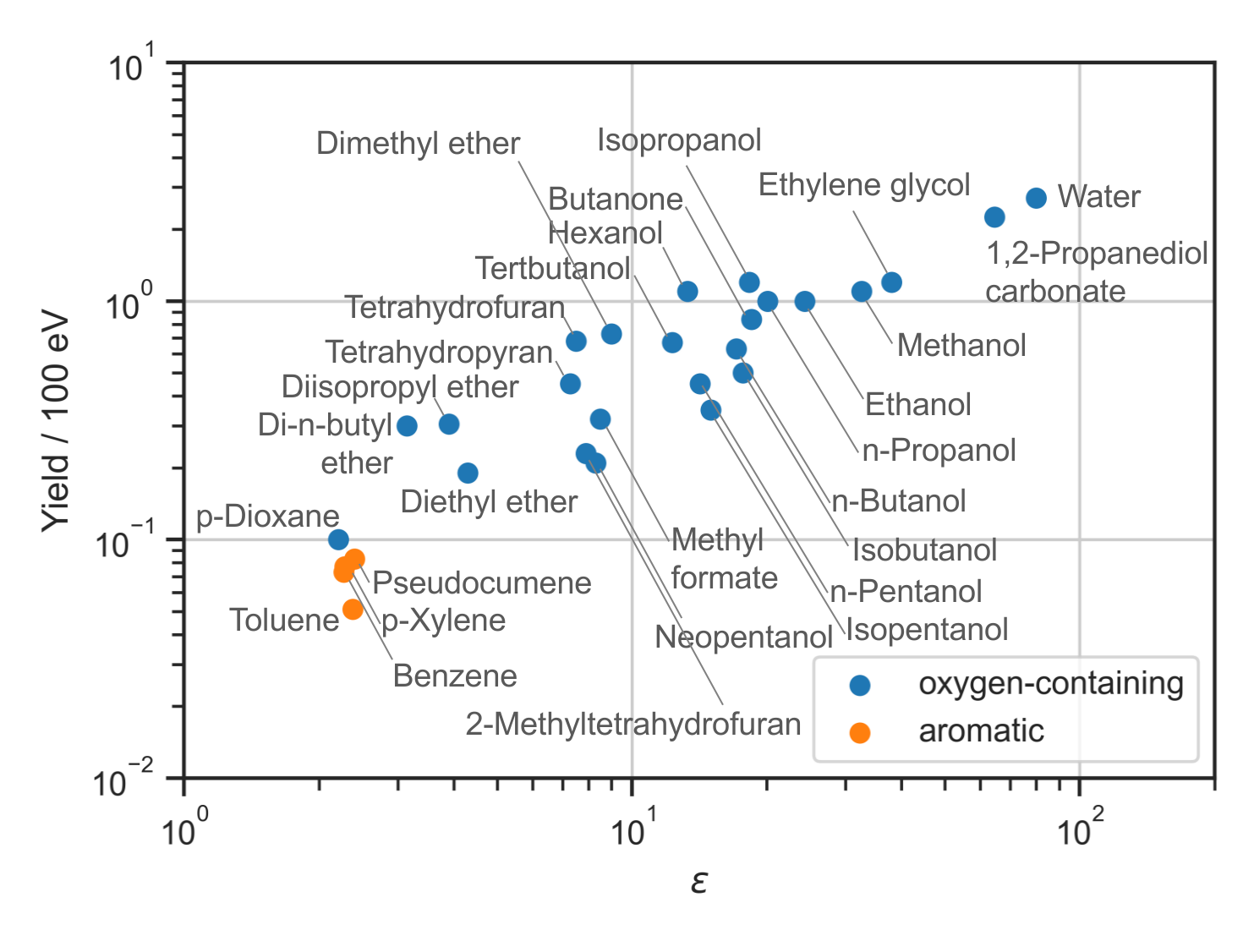}
    \caption{Free ion yield versus relative dielectric constant for water, four aromatic molecules, and some organic molecules containing the element oxygen, i.e.,~ether, alcohol, ketone, and ester~\cite{FreeIonYield}.}
    \label{fig:FreeIonYield}
\end{figure}

With the information compiled in Ref.~\cite{FreeIonYield}, 
the free ion yields of some organic molecules containing the element oxygen, i.e.,~ether, alcohol, ketone, and ester, are plotted with their relative dielectric constants in Fig.~\ref{fig:FreeIonYield}.
Water and four aromatic molecules, similar to LAB, are also plotted for comparison.
The data points lie in the diagonal region and are coarsely correlated.
For water, the free ion yield is 2.7/100~eV~\cite{FreeIonYield}.
For the four aromatic molecules, benzene, toluene, p-xylene, and pseudocumene, the free ion yield is less than 0.1/100 eV~\cite{FreeIonYield}.

In order to understand the importance of the combination of positive and negative ions, we compared the yield of free ions to the yield of excited states and to the yield of all ionization-generated electrons.

The total numbers of ionization electrons and excited states are estimated with a simulation tool.
Based on the Geant4 software, the Geant4-DNA package~\cite{G41, G42, G43, G4-DNA1,G4-DNA2,G4-DNA3,G4-DNA4,G4-DNA5} has been developed and extended with processes for the modeling of biological damage induced by ionising radiation at the DNA scale.
According to the introduction of Ref.~\cite{G4-DNA2}, the most recent sets of physics models  
(option 4 and option 6) enable more accurate simulation of stopping powers, dose point
kernels, and so on than the default set of models. 

KeV electrons are simulated in water with the physics model of option 4, and the numbers of all ionized electrons and excited states are counted, where no recombination is simulated.
The yields of total electrons and excited states are 4.4/100 eV and 2.3/100 eV, respectively.
We see that the free ion yield 2.7/100 eV is 
quite significant and takes about 2.7/(4.4+2.3) = 40\%.

With the same Geant4-DNA package, we adjusted several configurations for benzene simulation.
The molecular weight and density are modified as for benzene. 
The ionization, excitation, and elastic scattering cross-sections and related energy levels for water are replaced with those for benzene with the data given in Ref.~\cite{Xbenzene}.
The angular differential cross-section for ionization is set with Ref.~\cite{XDiff}. 
With the simulation of 1 keV electrons,
the yields of total ionization electrons and excited states are 5.6/100 eV and 6.0/100 eV, respectively.
The results are slightly higher than in water due to the smaller energy required for excitation and ionization. The ratio of free ion yield to the total yield of ionization electrons and excited states is $<$0.1/(5.6+6.0)=1\%. 
Compared to water, the four aromatic molecules have a very low free ion yield, and have more excited states contributing to the following scintillation process. 

\subsection{Electric dipole-dipole Energy transfer}
For the dipole-dipole process, in the language of quantum electrodynamics, the process is mediated by a virtual photon. The energy of the virtual photon is in the UV range, and its frequency is a few hundreds terahertz. The refractive indices of water and most organic liquids all drop to around 1, 
unless there are basic energy level structures of molecules causing strong absorption (Sec.~7.5 of Ref.~\cite{Jackson}). 
There is no significant difference between the different solutions.

\subsection{Light yield versus dielectric constant}
\begin{figure}[htpb]
    \centering
    \includegraphics[width=\linewidth]{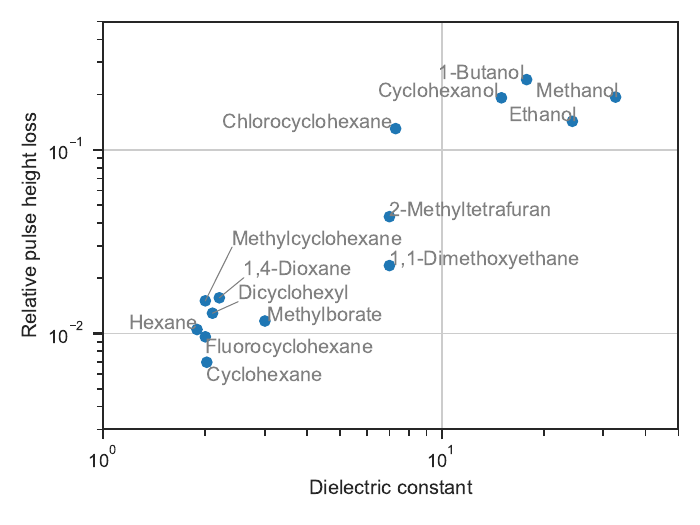}
    \caption{Light yield versus dielectric constant. The light yield results are compiled with the information in Ref.~\cite{Birks}.}
    \label{fig:yield-dielec}
\end{figure}

A plot is made based on the information given in the Fig.~8.2 of Ref.~\cite{Birks}.
There, 35 solvents were mixed with toluene containing 3~g/L PPO to check their quenching effect.
Relative pulse heights were measured with respect to the one without mixing.
The relative pulse height is an equivalent value of light yield, if the scintillation lifetime is not disturbed, which is the case of the following discussion.
To study the dependence on the dielectric constant, we selected 14 samples. 
The results with carbonyl are rejected, because carbonyl is known for accelerating intersystem crossing~\cite{Turro}.
The results with bromine and iodine are excluded, because heavy elements can cause stronger spin-orbit coupling and enhance intersystem crossing too.
Samples with benzene structures are excluded, because they are not pure quenchers.
Several alkanes, ethers, and alcohols are left and studied.
We extracted the pulse height at 5\% (by weight) mixing. The result with other mixing fractions, for example, 10\%, is similar.
Following the style of Fig.~\ref{fig:FreeIonYield}, a plot of the loss of pulse height versus dielectric constant is made, and is shown in Fig.~\ref{fig:yield-dielec}.
The number of samples is limited, but a coarse trend can also be observed. 
The polar structures play a role in quenching.

\section{Electric properties of TeBD and quenching}
\label{sec:TeBD}

\subsection{Synthesis of TeBD}
Following the procedure of tellurium loading of the SNO+ experiment~\cite{SNOPlusTeBD}, we synthesized a chemical with telluric acid, TeA, and 1,2-butanediol, BD, and the final product is called TeBD. 
The product is similar to the SNO+ publication, but not identical, and the details are explained below.
TeA with purity over 98\% was ordered from Shanghai Macklin Biochemical Technology Co., Ltd. BD with purity over 99\% was ordered from Beijing Konoscience Technology Ltd.
No further purification is applied. 
TeA aqueous solution and BD were mixed with a molar ratio of TeA to BD of 1:3. The mixture was heated and agitated continuously under vacuum to remove water at 70–80 $^\circ C$. The heating was applied until the product reached the solubility point in LAB.

\begin{figure}[htpb]
    \centering
    \includegraphics[width=\linewidth]{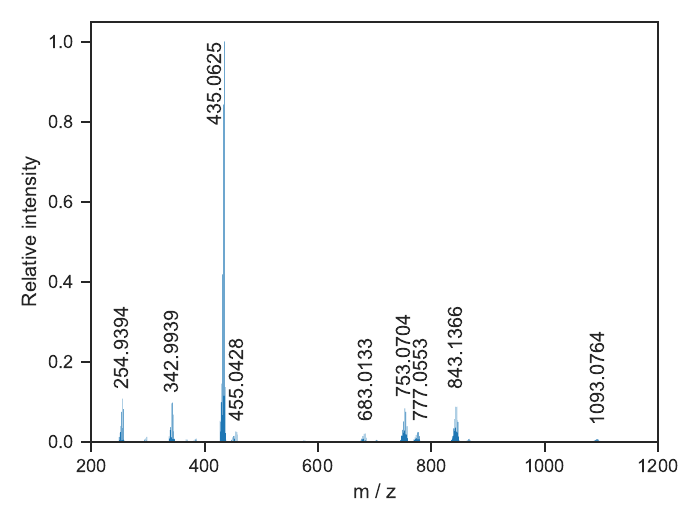}
    \caption{Positive ion mass spectrum for TeBD.}
    \label{fig:MassPosi}
\end{figure}

\begin{figure}[htpb]
    \centering
    \includegraphics[width=\linewidth]{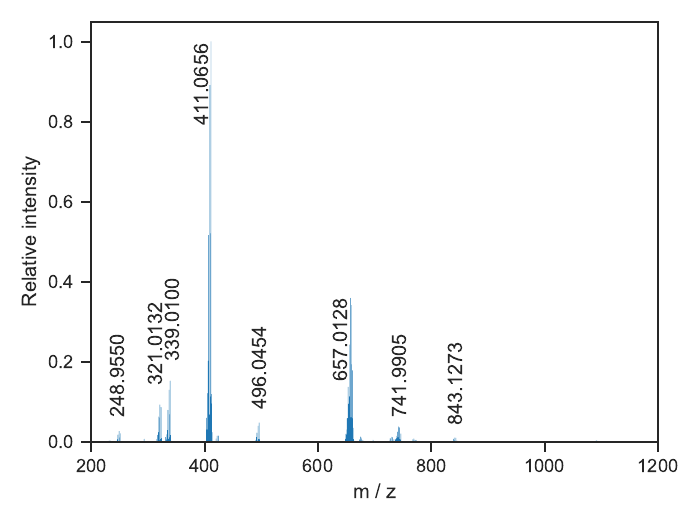}
    \caption{Negative ion mass spectrum for TeBD.}
    \label{fig:MassNega}
\end{figure}

Before the sample was analyzed with a LCMS-IT/TOF Mass Spectrometer, Shimadzu Japan, the sample was sealed and stored in our laboratory for a few months with inevitable contact with air and the vapor in the air. 
The analysis result is shown in Fig.~\ref{fig:MassPosi} and Fig.~\ref{fig:MassNega} for positive and negative modes, respectively. No additional two hours of heating were conducted before the mass spectrometer analysis, as in Ref.~\cite{SNOPlusTeBD}.

\begin{figure}[htpb]
    \centering
    \includegraphics[width=0.6\linewidth]{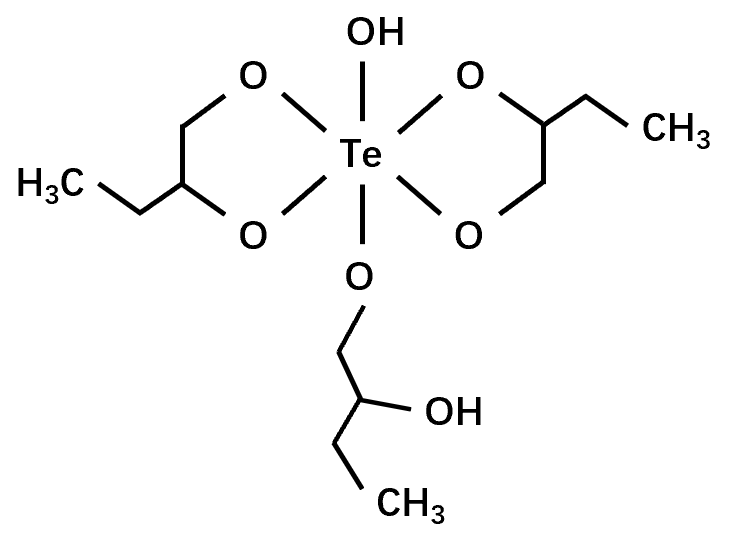}
    \caption{Molecule structure of TeBD at negative mode mass 411, which is the dominant synthesis product of this work.}
    \label{fig:StructureB}
\end{figure}

\begin{figure}[htpb]
    \centering
    \includegraphics[width=0.6\linewidth]{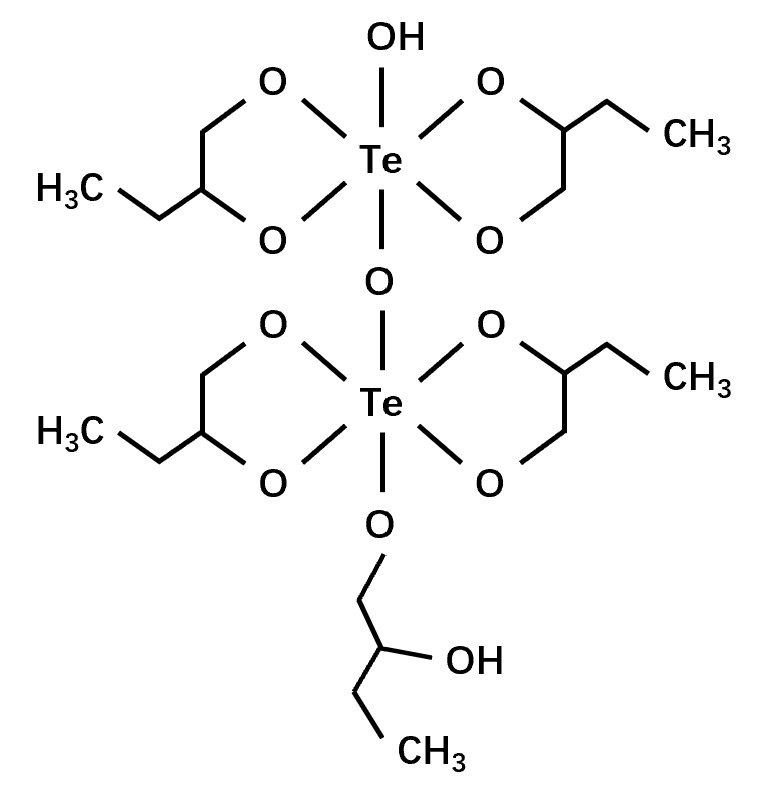}
    \caption{Molecule structure of TeBD at mass 729, which is the dominant synthesis product of Ref.~\cite{SNOPlusTeBD}.}
    \label{fig:StructureD}
\end{figure}

With the results of these analyses, the SNO+ TeBD product is basically reproduced.
The structure `b' in Fig.~4 of Ref.~\cite{SNOPlusTeBD} is visible on peak 435 in positive mode and peak 411 in the negative mode. 
This is the major product of our TeBD mixture product, and its structure, as suggested by Ref.~\cite{SNOPlusTeBD}, is replotted in Fig.~\ref{fig:StructureB}.
Note that in the positive mode, peak positions are possible to be shifted by adding the contribution of hydrogen or by sodium contamination, i.e.,~be shifted by their mass number 1 or 24, respectively, which is an intrinsic feature of the mass spectrometer.
The structure `d' in Fig.~4 of Ref.~\cite{SNOPlusTeBD}, which is the major product of SNO+ TeBD, can be seen on peak 753 in the positive mode with a mass of 729, if subtracting 24. 
Its structure is shown in Fig.~\ref{fig:StructureD}~\cite{SNOPlusTeBD}.
The structure `a' in Fig.~4 of Ref.~\cite{SNOPlusTeBD} is visible on peak 342 in the positive mode and peak 339 in the negative mode.
The structure `c' in Fig.~4 of Ref.~\cite{SNOPlusTeBD} is visible on peak 657 in the negative mode.
The structure `e' in Fig.~4 of Ref.~\cite{SNOPlusTeBD} is visible on peak 843 in the positive mode, if subtracting 24.
The major difference with Ref.~\cite{SNOPlusTeBD} is that 
our sample was not heated beyond the LAB solubility point and had contact with vapor, so there is much less oligomerisation in our sample.

\subsection{Electric property of TeBD}

\subsubsection{Polar groups in TeBD}
There are two hydroxyl groups in each TeBD compound, as shown in Fig.~\ref{fig:StructureB} and \ref{fig:StructureD}; one can expect they have similar effects on anions and cations as other alcohol substances.
These polar groups can slow down and prevent some of the anion-cation recombination back to the excited states for scintillation and can cause quenching. 

\subsubsection{Dielectric constant of TeBD}
The relative dielectric constant of TeBD in this work was measured with a dielectric constant and dielectric loss testing instrument, mode DHGTJS, produced by Shanghai Donghai Electric CO., LTD.

\begin{figure}[htpb]
    \centering
    \includegraphics[width=0.6\linewidth]{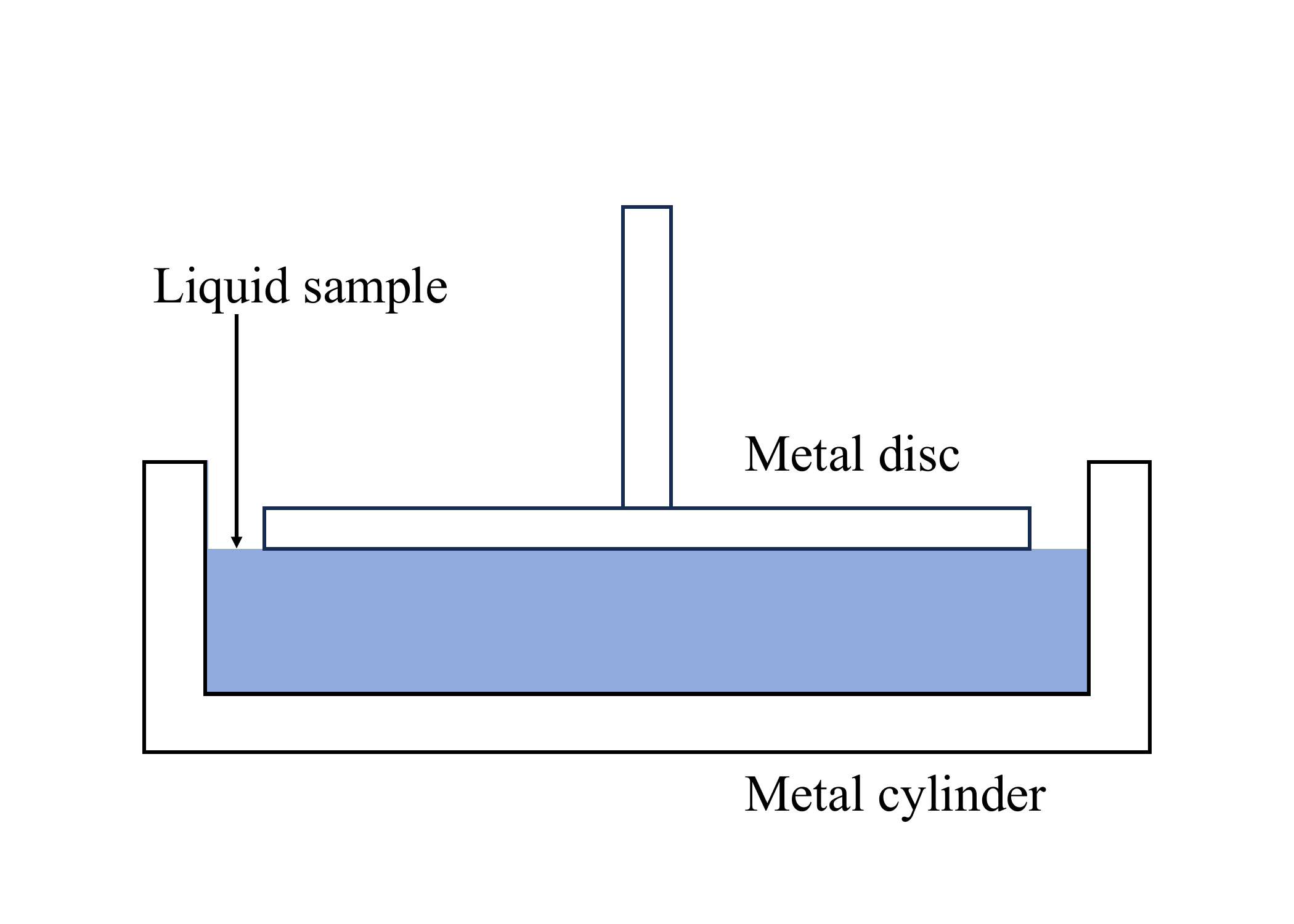}
    \caption{Liquid sample container of the dielectric constant measurement device..}
    \label{fig:container}
\end{figure}

\begin{figure}[htpb]
    \centering
    \includegraphics[width=0.6\linewidth]{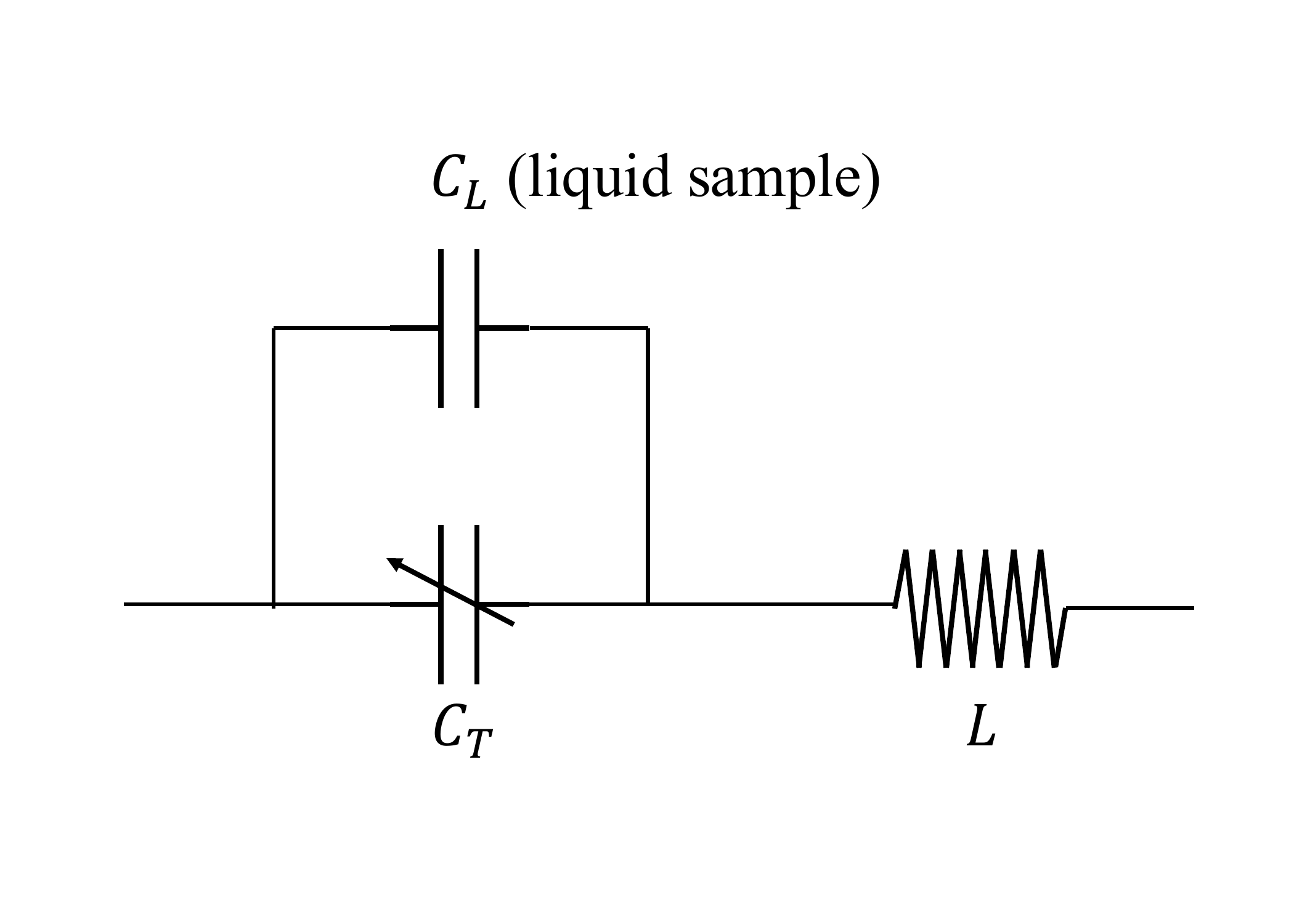}
    \caption{Equivalent circuit diagram of the dielectric constant measurement device.}
    \label{fig:circuit}
\end{figure}

The dielectric constant measurement device is briefly explained.
The device has a special metal cylinder container, as shown in Fig.~\ref{fig:container}. 
At a certain height, there is a circular metal disc. 
A liquid sample can be added to the cylinder container.
The cylinder, the liquid sample, and the disc form a capacitor.  
As shown in Fig.~\ref{fig:circuit}, the capacitor is 
connected in parallel with a tunable capacitor inside the device.
These two capacitors are connected with an inductor coil of the device, forming an LC circuit.
First, before a liquid sample is added, applying an alternating electric field of a fixed frequency, the capacitance of the variable capacitor is adjusted to reach a resonance of the LC circuit.
Then, the liquid sample is added, and the tunable capacitor is adjusted again to reach the LC resonance again.
The capacitance mediated by the liquid sample is known and the dielectric constant of the sample is deduced.


We first measured the relative dielectric constants of ethanol and cyclohexane at 1 MHz and 28~$^\circ C$. 
The frequency was limited by our instrument. The temperature was at room temperature. 
With several measurements, the results of ethanol are in the range of 24.3-25.7, 
and the literature reported result is 25.3 at 0 Hz and 20 $^\circ C$~\cite{CRCHandbook}.
The measured results of cyclohexane are in the range of 1.96-2.01, and the reported result is 2.02 at 0 Hz and $20~^\circ C$~\cite{CRCHandbook}.
These results basically verified the reliability of the instrument. 
With several measurements, the dielectric constant of LAB is in the range of 2.3-2.4. 

The result of the TeBD sample is measured to be $16\pm1$ at 1 MHz and 28~$^\circ C$.
The uncertainty is given by the difference of our measurements and the literature for ethanol and cyclohexane.

\section{Summary}
\label{sec:discussion}
In this paper, the energy deposition and scintillation processes in two component liquid scintillators are briefly reviewed, and the key features in anion-cation recombination are pointed out. 
Polar groups and a high dielectric constant of a liquid scintillator can slow down or prevent some of the anion-cation recombination and reduce scintillation. 
The case of TeBD is discussed and analyzed in this work. 
The relative dielectric constant of TeBD is measured to be $16\pm1$.
The polar groups and the related dielectric constant explain part of the reason for quenching.
Hopefully, some new chemicals and molecular structures for isotope loading can be identified in the future with fewer polar groups and a lower dielectric constant to reach a higher scintillation light yield.

\section{Acknowledgments}
This work is supported in part by
the National Natural Science Foundation of China (No.~12141503),
the Ministry of Science and Technology of China (No.~2022YFA1604704),
the Key Laboratory of Particle \& Radiation Imaging (Tsinghua University),
and the CAS Center for Excellence in Particle Physics (CCEPP).
Zhe Wang also likes to thank Minfang Yeh for helping me to digest many chemical concepts, and to thank Qing Wang and Zhicai Zhang.

\bibliographystyle{unsrtnat}
\bibliography{LSElectric}

\end{document}